\documentclass[useAMS,usenatbib]{mn2e}

\usepackage{times}
\usepackage{graphicx}

\newcommand{\beq}{\begin{equation}}
\newcommand{\eeq}{\end{equation}}
\newcommand{\lab}{\label}

\newcommand{\om}{\Omega_\mathrm{M 0}}
\newcommand{\ob}{\Omega_\mathrm{B 0}}
\newcommand{\oc}{\Omega_\mathrm{C 0}}
\newcommand{\ok}{\Omega_\mathrm{K 0}}
\newcommand{\ol}{\Omega_{\Lambda 0}}

\newcommand{\bfx}{\mathbf{x}}
\newcommand{\bfk}{\mathbf{k}}

\def\gs{\mathrel{\lower0.6ex\hbox{$\buildrel {\textstyle >}\over{\scriptstyle \sim}$}}}
\def\ls{\mathrel{\lower0.6ex\hbox{$\buildrel {\textstyle <}\over{\scriptstyle \sim}$}}}

\begin{document}

\title[Power spectra for modified gravity]{Imprints of deviations from
the gravitational inverse-square law on the power spectrum of mass
fluctuations}
\author[M. Sereno and J.~A. Peacock]{M. Sereno$^{1,2}$\thanks{E-mail:
sereno@physik.unizh.ch} and J.~A. Peacock$^{3}$
\\
$^{1}$Dipartimento di Scienze Fisiche, Universit\`{a} degli Studi di
Napoli `Federico II', Via Cinthia, Monte S. Angelo, 80126 Napoli,
Italy
\\
$^{2}$Institut f\"{u}r Theoretische Physik, Universit\"{a}t Z\"{u}rich,
Winterthurerstrasse 190, CH-8057 Z\"{u}rich , Switzerland
\\
$^{3}$Institute for Astronomy, University of Edinburgh, Royal
Observatory, Blackford Hill, Edinburgh EH9 3HJ, United Kingdom}

\date{May 18, 2006}

\maketitle

\begin{abstract}
Deviations from the gravitational inverse-square law would imprint
scale-dependent features on the power spectrum of mass density
fluctuations. We model such deviations as a Yukawa-like contribution
to the gravitational potential and discuss the growth function in a
mixed dark matter model with adiabatic initial conditions. Evolution
of perturbations is considered in general non-flat cosmological models
with a cosmological constant, and an analytical approximation for the
growth function is provided. The coupling between baryons and cold
dark matter across recombination is negligibly affected by modified
gravity physics if the proper cutoff length of the long-range
Yukawa-like force is $\gs 10~h^{-1}\mathrm{Mpc}$. Enhancement of
gravity affects the subsequent evolution, boosting large-scale power in
a way that resembles the effect of a lower matter density. This
phenomenon is almost perfectly degenerate in power-spectrum shape with the effect of a
background of massive neutrinos. Back-reaction on density growth from
a modified cosmic expansion rate should however also affect the normalization of
the power spectrum, with a shape distortion similar to the case of a
non-modified background.
\end{abstract}


\section{Introduction}

General relativity has passed many important tests up to the length and time
scales of the observable universe \citep{pee02,pee04}. The body of
evidence is quite impressive, considering the enormous extrapolation
from the empirical basis, but the gravitational inverse-square law and its
relativistic generalization are supported by high-precision tests from
measurements in the laboratory, the solar system and the binary pulsar
only up to scales of $\ls 10^{13}~\mathrm{cm}$ \citep{ald+al03}.

Despite the overall success of general relativity, non-standard
theories have received much recent attention, largely motivated by finding
alternative explanations for the `dark' sector of the universe. On one
hand, MOND proposals make gravity stronger on scales of galaxies to
explain flat rotation curves without dark matter \citep{mil83,san98}; on the
other hand, alternative theories aim to find a mechanism for
cosmic acceleration without dark energy, as a result of gravity
leaking into higher dimensions on scales comparable to the horizon \citep{dva+al00}.
Such radical modifications of gravitational physics may seem difficult
to test, but our ability to make precise observations of cosmological dynamics
on scales of $\gs 10~h^{-1}\mathrm{Mpc}$ is now good enough to permit
detailed tests of gravity theories even on these scales.

Examples of such work include \citet{wh+ko01}, who determined
constraints on the long-range properties of gravity from weak
gravitational lensing cosmic shear. 
Constraints on Newton's constant from the primordial abundances of light
elements were discussed by \citet{ume+al05}. 
The implications of MOND on large scale
structure in a Friedmann-Lema\^{\i}tre-Robertson-Walker (FLRW) universe
were discussed in \citet{nus02}. Recently, proposals have been made to
extend these theories to a general covariant framework \citep{bek04}
and evolution of perturbations in this theory has been considered
\citep{sko+al05}. Finally, large galaxy surveys have been
analysed looking for signatures in the power spectrum due to
deviations from the inverse-square law \citep{sea+al05,shi+al05}.
In the present paper, we shall concentrate on this last aspect.

We follow a number of recent authors, who have
considered additional contributions to the gravitational
potential in the form of Yukawa-like terms
\citep{wh+ko01,pee02,sea+al05,shi+al05}.
We consider the growth function and the power spectrum
of matter density fluctuations in these Yukawa-like models;
we extend previous work by removing a restriction to fixed
cosmological parameters and we also explore some possible
back-reactions from a modified cosmic expansion rate. We still adopt
the cold dark matter (CDM) model for structure formation with
adiabatic initial conditions, but we also consider variants due to hot
dark matter. The paper is structured as follows. In
Section~\ref{sec:ykw}, we introduce Yukawa-like contributions to the
gravitational potential and study the linear perturbation equation and
the growth function. In Section~\ref{sec:bar}, we give some insights
on the coupled growth of baryons and CDM. In Section~\ref{sec:pwr}, we
illustrate the effect of modified gravity on the power spectrum,
whereas Section~\ref{sec:mdm} is devoted to the interplay between
mixed dark matter and gravity physics. Growth of perturbations in a
model with modified cosmic expansion is addressed in Section
\ref{sec:csm}. Section~\ref{sec:con} contains some final
considerations.

\section{Linear perturbation equation}
\label{sec:ykw}

\begin{figure}
        \resizebox{\hsize}{!}{\includegraphics{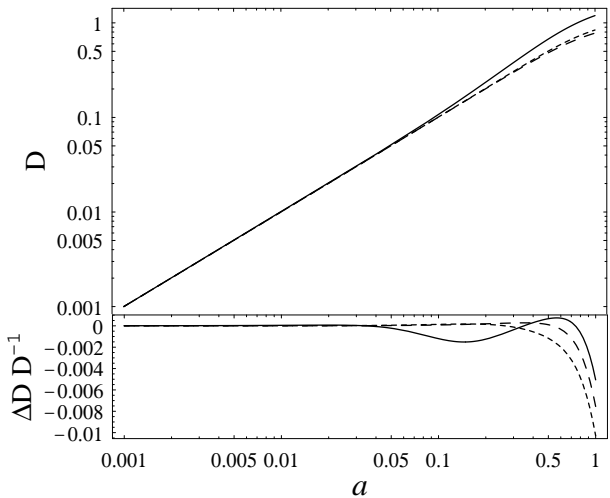}}
        \caption{Linear growth rate as a function of the scale factor $a$ in a
        flat $\Lambda$CDM model with $\om=0.25$ for Yukawa-model parameters
        $\Delta \alpha=0.5$ and $\lambda=10$~Mpc. Extra growth is effective at late
        times and on large scales. Upper panel: dependence on scale. 
	Extra growth is activated for $a \gs k \lambda$.
        The full, short-dashed
        and dashed lines are for $k=10^{-2},5{\times}10^{-2}$ and $10^{-1}\mathrm{Mpc}^{-1}$,
        respectively. Lower panel: fractional residuals between the above numerical
        results and the fitting formula.}
        \label{GrowthFunction_ykw}
\end{figure}

\begin{figure}
        \resizebox{\hsize}{!}{\includegraphics{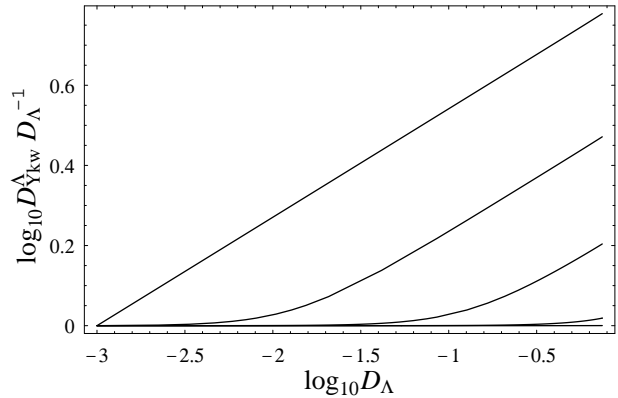}}
        \caption{Linear growth rate, normalized to the pure Newtonian case $D_\Lambda$
        ($\Delta \alpha=0$), as a function of $D_\Lambda$. The background cosmology is a
        flat $\Lambda$CDM model with $\om=0.25$, whereas the Yukawa-model parameters are
        fixed to $\Delta \alpha=0.5$ (i.e $p_{\Delta \alpha} \sim 0.27$) and $\lambda=10$~Mpc. The
        lines from above to below correspond to $k=10^{-6},10^{-3},10^{-2},10^{-1}$
        and $1~\mathrm{Mpc}^{-1}$, respectively. The growth function evolves as
        $D_\Lambda$ at very early times or small scales, and as $D_\Lambda^{1+p_{\Delta \alpha}}$
        when $a \gs \lambda k$.}
        \label{GrowthFunction_ykw_LargeScales}
\end{figure}

\begin{figure}
        \resizebox{\hsize}{!}{\includegraphics{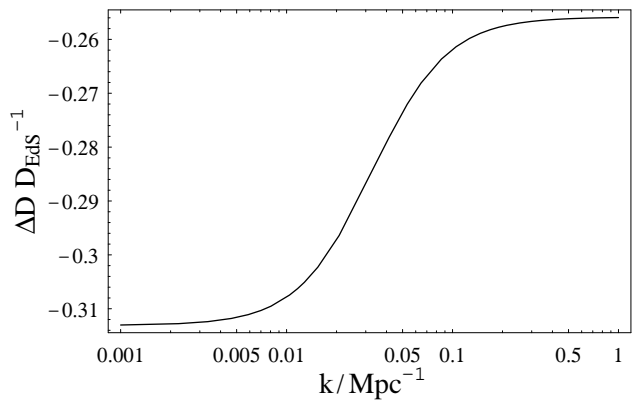}}
        \caption{Fractional deviation of the growth function, at the present time $a=1$,
        between an Einstein-de Sitter (EdS) model and a flat $\Lambda$CDM model with $\om=0.25$.
        The Yukawa-model parameters are $\Delta \alpha=0.5$ and $\lambda=20$~Mpc. The
        suppression due to a cosmological constant is scale dependent.}
        \label{DeltaGrowthFunction_ykw_EdStoLambda}
\end{figure}

The dynamics of linear perturbations can be easily extracted from a
Newtonian approach by writing down the fundamental equations of fluid
motion \citep[e.g.][chapter 15]{pea99}. Usually, acceleration is
related to mass density via Poisson's equation, but here we allow for
deviations. We model the weak field limit of the gravitational
potential, $\phi$, as a sum of a Newtonian potential and a Yukawa-like
contribution \citep[see][and references therein]{wh+ko01}:
\beq
\label{ykw1}
\phi  = \alpha_\mathrm{Nwt} \phi_\mathrm{Nwt} + \alpha_\mathrm{Ykw} \phi_\mathrm{Ykw}.
\eeq
The factors $\alpha_\mathrm{Nwt}$ and $\alpha_\mathrm{Ykw}$ account
for deviations of the effective gravitational coupling constant from
$G_\mathrm{N}$, the value actually observed on small scales.
$\phi_\mathrm{Nwt}$ is the solution of Poisson's equation, whereas
$\phi_\mathrm{Ykw}$, which describes forces mediated by massive
particles, is the solution of the modified Poisson equation; in
comoving coordinates, this can be written as
\beq
\label{ykw2}
\frac{\nabla^2 \phi_\mathrm{Ykw} }{a^2} - \frac{\phi_\mathrm{Ykw} }{\lambda^2} =
4 \pi G_\mathrm{N} \rho_\mathrm{M} \delta (\bfx, t),
\eeq
with $a$ the cosmic scale factor, $\rho_\mathrm{M}$ the total matter
density, $\delta$ the mass density contrast and $\lambda$ a proper
length cutoff. The potential in Eq.~(\ref{ykw1}) is $\propto 1/x$ both
on small scales ($ x \ll
\lambda/a$), with an effective coupling constant $G=(
\alpha_\mathrm{Nwt} +\alpha_\mathrm{Ykw})G_\mathrm{N}$, and on large scales,
with $G_\infty = \alpha_\mathrm{Nwt} G_\mathrm{N}$. Putting
\beq
\alpha_\mathrm{Nwt} \equiv 1+ \Delta \alpha
\eeq
and $\alpha_\mathrm{Ykw} = -
\Delta \alpha$, $\phi$ is in numerical agreement with results on the
scale of the solar system, whereas $G_\infty =(1+
\Delta \alpha) G_\mathrm{N}$. For $\Delta \alpha >0(<0)$,
gravity is enhanced (suppressed) on large scales. The potential in
Eq.~(\ref{ykw1}) can be derived in a relativistic gravity model that
obeys the equivalence principle \citep{zh+ne94}. Although the
Yukawa-like form is quite specific, all long-range
deviations can be characterised empirically by an amplitude, $\Delta \alpha$, and
a length scale, $\lambda$.

The basic equation for the linear evolution of the Fourier component
$\delta_\bfk$ of the density fluctuation, with comoving wave-number
$\bfk$, is
\beq
\lab{lin1}
\ddot{\delta}_\bfk + 2 H(t)\dot{\delta}_\bfk = -\frac{k^2 \phi_\bfk}{a^2},
\eeq
where dot denotes time derivative and $H \equiv \dot{a}/a$ is the time
dependent Hubble term. In Fourier space, the potential in
Eq.~(\ref{ykw1}) is \citep{wh+ko01,sea+al05}
\beq
\lab{lin2}
\phi_\bfk = -\frac{4 \pi G_\mathrm{N} a^2 \rho_\mathrm{M}
\delta_\bfk }{k^2} {\cal Y}(a/a_\lambda ;\alpha_\mathrm{Nwt},\alpha_\mathrm{Ykw}),
\eeq
with
\begin{eqnarray}
{\cal Y} & \equiv  & \alpha_\mathrm{Nwt}+\frac{
\alpha_\mathrm{Ykw} } {1+\left(a/a_\lambda \right)^2} \\
& = & 1+\Delta \alpha \frac{
\left( a/a_\lambda \right)^2 }{1+\left(a/a_\lambda \right)^2}
\end{eqnarray}
and $a_\lambda \equiv  \lambda k$. From Eqs.~(\ref{lin1},~\ref{lin2}),
we obtain \citep{sea+al05,shi+al05}
\beq
\label{lin2bis}
\ddot{\delta}_\bfk + 2 H \dot{\delta}_\bfk+
\frac{3}{2} \frac{\Omega_\mathrm{M0} H_0^2}{a^3}
{\cal Y}(a/a_\lambda ;\alpha_\mathrm{Nwt},\alpha_\mathrm{Ykw})
\delta_\bfk = 0,
\eeq
where $\Omega_\mathrm{M0}$ is the present density of total matter in
units of the critical one and $H_0$ is the Hubble constant,
parameterized in the following as $100~h~\mathrm{km~s^{-1}Mpc^{-1}}$.
A deviation from the inverse-square law thus causes the rate of growth
of the Fourier amplitudes to depend on wavelength. The inclusion of a
Yukawa-like contribution imprints the scale $\lambda/a_0$ on the
power spectrum. Let us examine the growth between an initial epoch,
$a_\mathrm{in}$, and the actual time, $a_0$, in more detail. On a very
large scale, i.e. $k \ll a_\mathrm{in} /
\lambda$ and ${\cal Y } \sim 1+ \Delta \alpha$, modes always
experience the modified growth due to the effective
gravitational constant. Intermediate scales, initially below the
maximal comoving cutoff length $\lambda/a_\mathrm{in}$, will cross
the scale for modified growth at some time $ a_\mathrm{in}\ls
a_\lambda\ls a_0 $. Finally, very small scales $k \gg a_0~\lambda$ will
evolve in a Newtonian potential. As a result, the growth of
fluctuations is no longer independent of scale, even at low redshift.
Some examples of the growth function are plotted in
Fig.~\ref{GrowthFunction_ykw}.

Equation~(\ref{lin1}), with $\phi$ given as in Eq.~(\ref{lin2}), has
been solved analytically for an Einstein-de Sitter model (EdS, $\om=1,
\ok=0$) in terms of hypergeometric functions in
\citet{shi+al05}.\footnote{We refer to Section~\ref{sec:csm} for
considerations about modified cosmic expansion rate and adopt here a
standard FLRW framework.} Here, we want to examine approximate
solutions in the general case with cosmological constant. Let us begin
with the EdS model. A Yukawa-like contribution with $\Delta
\alpha >0 (<0)$ enhances (suppresses) the growth of fluctuations on
scales sufficiently larger than $\lambda$. The growth law becomes
\beq
\label{eds1}
\delta \propto a^{1+p_{\Delta \alpha}},\ p_{\Delta \alpha} = \frac{ {\sqrt{25 + 24\Delta \alpha}} -5}{4} .
\eeq
The growth function, $D$, of the CDM+baryons density fluctuations can
be approximated by smoothly interpolating between the analytic
results. We propose, for the growing mode,
\beq
\label{eds2}
D_\mathrm{Ykw}^\mathrm{EdS}(a;\Delta \alpha,\lambda) =
\delta_\mathrm{in} \frac{a}{a_\mathrm{in}}
\left\{ 1 +\left[ \frac{ a/a_\mathrm{in} }{1+ c_2 a_\lambda / a_\mathrm{in} }
\right]^{c_1} \right\}^{p_{\Delta \alpha}/c_1}
\eeq
where the normalization has been chosen to be $D=\delta_\mathrm{in}
a/a_\mathrm{in}$ at early times $a \sim a_\mathrm{in}$ and the
coefficients $c_1$ and $c_2$ represent a fit to numerical evolution.
For $a_\mathrm{in}=10^{-3}$, we get
\begin{eqnarray}
\lab{eds3}
c_1 & = & 1.94088 \\ c_2 & = & 1.42350 - 0.170668 \Delta \alpha
\end{eqnarray}
The fitting formula works quite well, for
$10^{-2}\ls~k/(h~\mathrm{Mpc}^{-1})\ls~1$, for the parameter range $
-0.25\ls \Delta \alpha \leq 1$ and $
2 \ls \lambda/(h^{-1}\mathrm{Mpc})\ls 100$, with fractional residuals
always under 1\%.

The growth function changes at late times when $\om~\neq~1$ since,
after matter ceases to dominate the expansion rate, fluctuation growth
halts. This must be considered together with the late-time effect due
to deviations from the inverse-square law since the comoving scale for
modified growth, $\lambda/a$, moves to smaller values with increasing
time. It is easy to account for both of these effects. Let $D_\Lambda$
denote the growth function for a model with non-zero curvature and
cosmological constant but in the absence of Yukawa-like terms. Then,
by the replacement of $a/a_\mathrm{in}$ with $D_\Lambda$ in
Eq.~(\ref{eds2}), we can approximate all the dependence of the
scale-dependent growth function on time, curvature and cosmological
constant in presence of a modified gravitational potential. A useful
approximation for $D_\Lambda$ can be found in \citet{car+al92}. In the
lower panel of Fig.~\ref{GrowthFunction_ykw}, we compare the numerical
evaluation to the fitting formula. The accuracy is quite high, with
performance better than 1\% for a parameter space similar to the EdS
case. The fitting formula approximates the asymptotic limit on very
large scales or late times, i.e $a \gs a_\lambda$, as
$D_\Lambda^{1+p_{\Delta\alpha} }$. This is a natural generalization of
the $a^{1+p_{\Delta\alpha} }$ form that applies for the EdS model.
Although the result is no longer exact if $\ol \neq 0$, it is very
accurate in practice. In Fig.~\ref{GrowthFunction_ykw_LargeScales}, we
can see how the linear growth index changes from the usual Newtonian
trend to an enhanced index for decreasing wave-numbers and decreasing
redshifts and how well the $D_\Lambda^{1+p_{\Delta\alpha} }$
approximation works at late times.

In the presence of a Yukawa-like term, the suppression due to a
cosmological constant with respect to the EdS model is
most effective at large scales, in contrast to the Newtonian
case, where the suppression is scale-independent. Let us consider the
case $\Delta
\alpha>0$. Whereas a Yukawa-like contribution is effective at late times and
on large scales, extra growth is inhibited by a cosmological constant
when matter becomes sub-dominant in the dynamics. In
Fig.~\ref{DeltaGrowthFunction_ykw_EdStoLambda}, we plot the fractional
deviation in the growth function between an EdS cosmology and a flat
$\Lambda$CDM model.

The hypotheses used above are standard in modelling the effect of a
Yukawa-like contribution to the gravitational potential on the growth
function \citep{sea+al05,shi+al05}. In the next sections, we will
relax some assumptions and extend the analysis to mixed dark matter
models.

\section{Baryons}
\label{sec:bar}

Baryons imprint characteristic features into the power spectrum as the
direct result of small density fluctuations in the early universe
prior to recombination, when they are tightly coupled with the photons
and share in the same pressure-induced oscillations \citep{ei+hu98}.
As far as the Yukawa-like potential affects all kinds of matter in the
same way, the modified coupled equations for perturbation growth in a
mixture of baryons and CDM can be written as
\begin{eqnarray}
\label{bar1}
\lefteqn{
\left( \begin{array}{c}
\ddot{\delta_\mathrm{B}} \\ \ddot{\delta_\mathrm{C}}
\end{array} \right)  + 2 H
\left( \begin{array}{c} \dot{\delta_\mathrm{B}} \\ \dot{\delta_\mathrm{C}} \end{array}
\right) =
}
\\
& & 4 \pi G \frac{\rho_\mathrm{M}}{\om} {\cal Y}
\left(
\begin{array}{cc}
\ob -k^2/(k_\mathrm{J}^2 {\cal Y} )  & \oc \\
\ob & \oc  \end{array}
\right)
\left( \begin{array}{c} \delta_\mathrm{B} \\ \delta_\mathrm{C}  \end{array} \right) \nonumber
\end{eqnarray}
where $\delta_\mathrm{B}$ and $\delta_\mathrm{C}$ are the density
fluctuations in baryons and CDM, respectively, $k_\mathrm{J} \equiv a
\sqrt{4
\pi G_\mathrm{N} \rho_\mathrm{M}}/c_\mathrm{s}$ is the comoving Jeans wave-number
in a Newtonian potential and $c_\mathrm{s}$ is the sound speed. On a
small scale, baryon pressure causes the growth rates in baryons and
CDM to differ. Gravity-driven growth is not allowed for baryons above
a critical wave-number, if $k/(k_\mathrm{J} \sqrt{\cal Y}) \gg 1 $. This
condition can be re-written for small values of $\Delta \alpha$ as
\beq
\label{bar2}
k \gg k_\mathrm{J} \left( 1 + \frac{\Delta \alpha /2 }{1 + (k_\mathrm{J}
\lambda/a)^2 } \right)+ {\cal O}(\Delta \alpha^2).
\eeq
A positive value of $\Delta \alpha $ pushes the domain of predominance
of pressure over gravity to smaller lengths. When pressure overcomes
gravity, a Yukawa-like term does not affect the dynamics of baryons.
As usual, their main behaviour will be oscillatory, with slowly
declining sound waves described by the WKB solution
\citep[e.g.][equation (15.60)]{pea99}.

Either on large scales or after the Compton drag epoch, when the
sound speed drops by a large factor, pressure effects are negligible.
Solutions of Eq.~(\ref{bar1}) are simple since the matrix in the
driving term has time independent eigenvectors: $(1,1)$ and
$(\oc,-\ob)$. Hence, a coupled perturbation will be quickly dominated
by the fastest-growing mode with $\delta_\mathrm{B}
= \delta_\mathrm{C}$ and baryons quickly catch up with the CDM.

Since we have been considering a proper length $
\lambda\gs~10~h^{-1}\mathrm{Mpc}$, the interaction of
matter and radiation at decoupling
is not affected by modified gravity:
the cutoff at this time is $\gs 10~h^{-1}\mathrm{Gpc}$ in comoving units, 
so its effects appear on a scale too large to be observed.
Prior to the drag epoch, the comoving baryonic Jeans scale,
$\lambda_\mathrm{J} \equiv 2\pi/k_\mathrm{J}$, tracks the sound
horizon, which is of the order of $100~h^{-1}\mathrm{Mpc}$ at $a \sim
10^{-3}$. Hence, for a proper cutoff length
$\lambda\gs~10~h^{-1}\mathrm{Mpc}$, the effects of a Yukawa-like term
on the effective Jeans wave-number, Eq.~(\ref{bar2}), are really
negligible. After the drag epoch, the oscillatory mode of baryons
matches onto a mixture of pressure-free growing and decaying modes. Such
modes are given by the solutions of the perturbation equation
discussed in Section \ref{sec:ykw} for a $\om = 1 $ matter dominated
universe. Because the form of the WKB solution is preserved and
because the modes after recombination are affected by modified gravity
only at later times, velocity overshoot is still produced with the
output after the drag epoch being maximised when the wave consists of
a pure velocity perturbation.

Although the position of the acoustic peaks is nearly unaffected, later
evolution can still affect the scales of baryonic oscillations
for small values of $\lambda$ and large values of $\Delta
\alpha$. A Yukawa-like term can alter the height of peaks, with the first
peak enhanced with respect to the second one when $\Delta \alpha >0$.
Baryonic acoustic oscillations have been proposed as probes of dark
energy \citep{bl+ga03,se+ei03,mat04}, but most of the attention is on
position rather than height, so that the effect of modified gravity is
negligible at the present status of accuracy of galaxy redshift
surveys \citep{col+al05}.

\section{Power Spectrum}
\label{sec:pwr}

\begin{figure}
        \resizebox{\hsize}{!}{\includegraphics{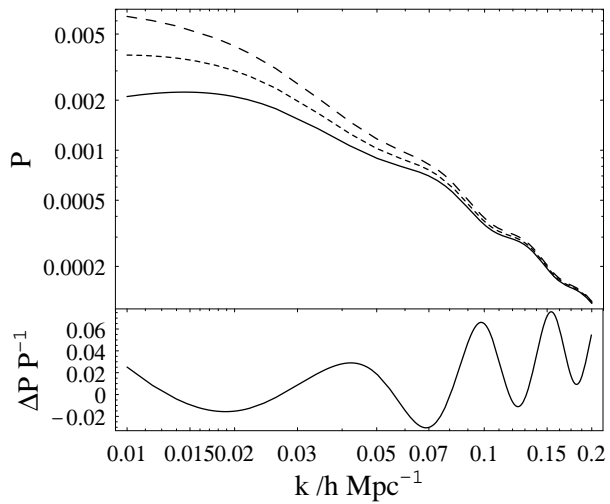}}
        \caption{Effect of a Yukawa-like term on the power spectrum.
        Upper panel: the full, dashed and long-dashed lines are for
         $\Delta \alpha=0,0.3$ and $0.6$ respectively, with
         $\lambda$ being fixed at $10~h^{-1}\mathrm{Mpc}$ in a flat
         $\Lambda$CDM model with $h=0.7,\om=0.25,\ob=0.05$
         and $n=1$. Units of the power spectrum are arbitrary. An enhancement
         of gravity ($\Delta \alpha>0$) boosts large-scale power.
         Lower panel: fractional residuals between the pure Newtonian
         case of the upper panel and a re-normalized power spectrum
         in a flat model with $h=0.7,\om=0.30,\ob=0.05, n=1$ and
         Yukawa-model parameters $\Delta \alpha=0.19, \lambda =5~h^{-1}\mathrm{Mpc} $.
         The effect of enhanced gravity is similar to that of lower matter density.}
        \label{P_k_ykw}
\end{figure}

The linear power spectrum can be constructed from the transfer
function as
\beq
\lab{pwr1}
P(k,a) \propto k^n T^2 (k)
\left[ \frac{D(z,a)}{D(z,a_\mathrm{in})} \right]^2,
\eeq
where $T(k)$ is the matter transfer function at the drag epoch, and
$n$ is the initial power spectrum index, equal to 1 for a scale
invariant spectrum. As discussed in the previous section, we can
assume that standard theory provides the transfer function at the last
scattering surface. In what follows, we will use the fitting formulae
for $T(k)$ in \citet{ei+hu98,ei+hu99}.

A positive long-range contribution to the gravitational potential
enhances clustering on scales larger than $\lambda$. The shape
distortion of the power spectrum can be quite significant: see
Fig.~\ref{P_k_ykw}. A Yukawa-like contribution with $\Delta
\alpha >0$ changes the spectrum in a way that resembles a lower matter
density, with a near degeneracy between the pair $\left\{
\Delta \alpha, \lambda \right\}$ and $\om$. An illustrative example is
contained in the lower panel of Fig.~\ref{P_k_ykw}. Larger fractional
residuals appear at the scale of the acoustic peaks, since in our example
we have varied $\om$ but kept $\ob$ fixed. As a matter of fact, the
location of baryonic oscillations is nearly independent of $\Delta
\alpha$ and could break the compensation. As well as a change in
density parameters, a Yukawa-like contribution can mimic a variation
in the primaeval spectral index, with values of $n <1$ nearly
degenerate with $\Delta \alpha >0$, over a large range of
wave-numbers.

The degeneracy between matter density and Yukawa parameters could be
quite significant when determining cosmological parameters, since
power-spectrum analyses provide one of the most important direct
constraints on the value of $\om$, whereas other methods that are
nearly independent of a Yukawa-like contribution, i.e. methods based
on the cosmic microwave background radiation or supernovae of type Ia,
are mostly sensitive to other combinations of cosmological parameters.

\section{Mixed dark matter}
\label{sec:mdm}

\begin{figure}
        \resizebox{\hsize}{!}{\includegraphics{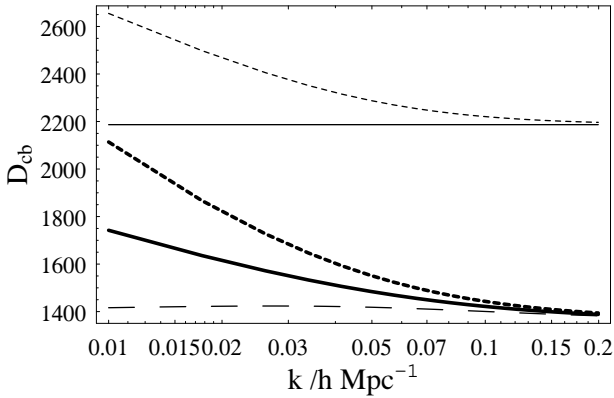}}
        \caption{Scale-dependent growth function in a MDM cosmology, evaluated
        at the present epoch. We consider a flat $\Lambda$CDM model with $h=0.7, \om=0.25$.
        Full lines represent Newtonian cases ($\Delta \alpha=0$);
        thick lines include neutrinos (3 species and $f_\nu=0.10$); short-dashed
        lines account for a Yukawa-like term with $\Delta \alpha=0.20$ and $\lambda =
        12~h^{-1}\mathrm{Mpc}$. Scale dependent features due to massive neutrinos are
        nearly degenerate with modified gravity. The long-dashed curve stands for
        the growth function with $\Delta \alpha=-0.20$, $\lambda = 12~h^{-1}\mathrm{Mpc}$,
        3 neutrino species and $f_\nu=0.05$. Effects due to neutrinos can be cancelled
        by a Yukawa-like term with $\Delta \alpha< 0$}
        \label{GrowthFunction_mdm_ykw}
\end{figure}

\begin{figure}
        \resizebox{\hsize}{!}{\includegraphics{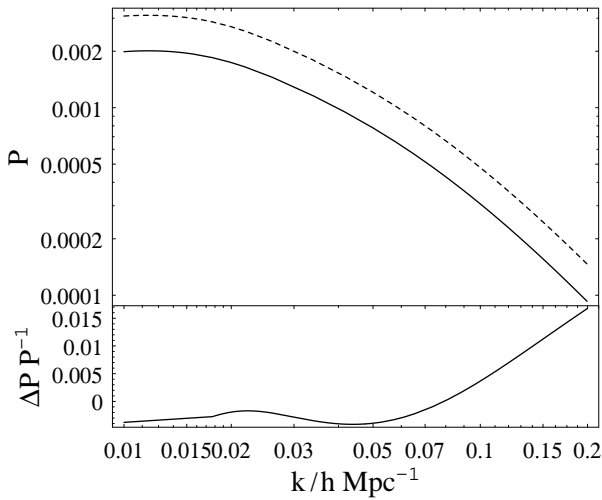}}
        \caption{ The effect of massive neutrinos on the shape of the power spectrum
        is nearly degenerate to that of a Yukawa-like contribution. We consider a
        flat $\Lambda$CDM model with $h=0.7,\om=0.25,
        \ob/\om = 0.15$ and $n=1$. Upper panel: the full line represents
        the power spectrum of the density weighted CDM, baryon and neutrino perturbations
        for $f_\nu=0.05$, 3 neutrino species and Newtonian potential ($\Delta \alpha=0$);
        the dashed line stands for $\Delta \alpha=0.19$ and $\lambda = 10~h^{-1}\mathrm{Mpc} $
        without neutrinos. The units for the power spectrum are arbitrary.
        Lower panel: fractional residuals after re-scaling.}
        \label{P_k_mdm_ykw}
\end{figure}

The introduction of massive neutrinos or other forms of hot dark
matter couples the spatial and temporal behaviour of perturbation
growth \citep{ma96,hu+ei98,ei+hu99}. On scales smaller than the
free-streaming length, neutrinos do not trace CDM and baryon
perturbations, which in turn grow more slowly because of the reduction
of the gravitational source. As neutrinos slow down and their
free-streaming scale shrinks to even smaller scales with increasing
time, the transfer function acquires a non-trivial time dependence.

As has been discussed in \citet{hu+ei98,ei+hu99}, the transfer
function in a mixed dark matter (MDM) cosmology with CDM, baryons and
massive neutrinos can be decomposed into a scale-dependent growth
function that incorporates all post-recombination effects, and a time
independent function that represents the perturbations around
recombination. The late-time evolution of density-weighted CDM and
baryons perturbations can be accurately fitted by \citep{hu+ei98}
\beq
\label{mdm1}
D_\mathrm{cb}(y,k) = \left\{ 1+ \left[
\frac{D_\Lambda(y)}{1+y_\mathrm{fs}(k)}
\right]^{0.7} \right\}^{p_\mathrm{cb}/0.7} D_\Lambda(y)^{1-p_\mathrm{cb}},
\eeq
where $D_\Lambda$ is the growth function in absence of neutrino
free-streaming, $y \equiv a/a_\mathrm{eq}$, with $a_\mathrm{eq}$ the
epoch of matter-radiation equality, $y_\mathrm{fs}$ is the
free-streaming epoch as a function of scale, and $p_\mathrm{cb} \equiv
\left(5-\sqrt{1+24f_\mathrm{cb}}
\right)/4$ with $f_\mathrm{cb}$ the matter fraction in CDM+baryons. The
fitting formula in Eq.~(\ref{mdm1}) interpolates across
$y_\mathrm{fs}$ between the suppressed logarithmic growth rate at
small scales, which is $1-p_\mathrm{cb}$ in a EdS model, and the usual
growth rate at sufficiently large scales. An analogous formula for the
growth rate for the CDM+baryon+neutrino perturbations can be found in
\citet{hu+ei98}. If we replace $D_\Lambda$ with the growth function
$D_\mathrm{Ykw}^\Lambda$, discussed in Sec.\ref{sec:ykw},
$D_\mathrm{cb}$ will contain all the dependence of the transfer
function on time, curvature, cosmological constant and Yukawa-like
contribution in presence of neutrino free-streaming\footnote{Note that
the growth functions in \citet{hu+ei98} have been normalized to
$D=a/a_\mathrm{eq}$ at early time, so that $D_\mathrm{Ykw}$ must be
conveniently re-scaled.}. An amplitude $\Delta \alpha >0$ has a
similar effect to that of a non-zero neutrino mass, both reinforcing
large-scale evolution. Some examples of growth functions are plotted
in Fig.~\ref{GrowthFunction_mdm_ykw}. Scale-dependent features due to
neutrinos can be almost perfectly cancelled by a Yukawa-like term.

It turns out that features in the power spectrum due to the transition
between free-streaming and infall of massive neutrinos have much in
common with imprints induced by enhanced gravity: see
Fig.~\ref{P_k_mdm_ykw}, where the fitting formulae from
\citet{ei+hu99}, which do not account for baryonic oscillations, have
been used. An enhancement of gravity ($\Delta \alpha >0$) can distort
the shape of the power spectrum in nearly the same way as a non-zero
neutrino mass, both boosting large-scale power. Even at low redshifts,
some scales can be still either in the free-streaming regime or below
the $\lambda$-cutoff, so there is an interplay between the two
effects. A true deviation from the inverse-square law can be
interpreted as a significant neutrino fraction $f_\nu$ of the total
matter density: see the lower panel of Fig.~\ref{P_k_mdm_ykw}. In a
low density flat model with three nearly degenerate neutrino species
and $f_\nu \ls 0.10$, an adequate approximation for such a degeneracy,
is, if $\lambda \ls 50~h^{-1}\mathrm{Mpc}$,
\beq
(f_\nu)_\mathrm{apparent} \simeq \left[ 0.312 - 0.252
\left( \frac{\lambda}{50}\right)\right] ( \Delta
\alpha )_\mathrm{true}
\eeq
with the proper length $\lambda$ in units of $h^{-1}$~Mpc. For a smaller
cutoff proper length, a larger deviation $\Delta \alpha$ is needed to
mimic the effect of massive neutrinos. This compensation works up to
the scales where nonlinear or redshift-space effects begin to affect
the shape of the power spectrum, i.e. for $k \ls
0.1~h~\mathrm{Mpc}^{-1}$ \citep{mei+al99}.

The fractional contribution of neutrinos to the total mass density has
been derived from the 2dFGRS data \citep{elg+al02}. Adding prior
information from independent cosmological probes, \citet{elg+al02}
found $f_\nu <0.13$ (at 95\% confidence), i.e an upper bound on the
total neutrino mass $m_\nu<1.8$~eV. \citet{spe+al03} combined 2dFGRS
data with measurements from WMAP and obtained a tight limit $m_\nu
<0.69$~eV. These results are strongly dependent on the assumed priors
\citep{el+la03} and could be significantly affected by a long-range
deviation from the inverse-square law. Since a long-range enhancement
could equally lead to extra large-scale power, the upper bound on the
neutrino mass would be over-(under)estimated if $\Delta
\alpha >0 (<0)$.

This degeneracy applies only at one epoch, and the evolution of $P(k)$
will be different in the two cases. Although neutrinos mainly affect
the growth function, they also contribute a residual suppression of
power in the transfer function on small scales at the drag epoch. On
the other hand, a Yukawa-like contribution, which is characterized by
a proper length cutoff, is only effective at later times.

\section{Modified cosmic expansion}
\label{sec:csm}

\begin{figure}
        \resizebox{\hsize}{!}{\includegraphics{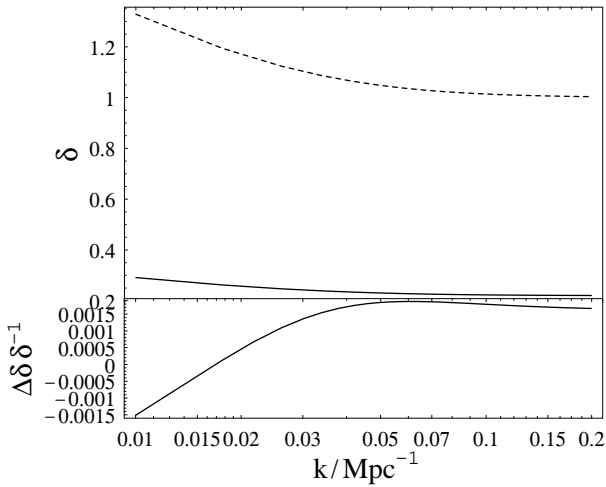}}
        \caption{The growth function, evaluated at the present epoch, in a model with
        modified background expansion. The shape of scale-dependent features due to a
        Yukawa-term are negligibly affected by a modified cosmic rate.
        Upper panel: the full line stands for a modified EdS
        model with $\Delta \alpha = 0.5$ and $\lambda =20~\mathrm{Mpc}$; the short-dashed
        line represents the growth function in a non-modified EdS model with
        $\Delta \alpha = 0.4$ and $\lambda =21~\mathrm{Mpc}$. Lower panel: fractional residuals after re-scaling.}
        \label{GrowthFunction_ykw_mod}
\end{figure}

In the previous sections we have followed the usual approach by
assuming that a modified gravitational potential only affects
fluctuations while leaving the background cosmology intact
\citep{wh+ko01,nus02,sea+al05,shi+al05}. We now want to address the
effect of a perturbed cosmological expansion rate, due to modified
gravity physics, on the dynamics of large-scale structure. In fact,
Lagrangian-based theories of modified Newtonian gravity, in which the
FLRW background cosmology still holds in the absence of fluctuations,
have been proposed \citep{san01}, but such frameworks are still
controversial. Possible cosmological implications of new long-range
gravitational interactions have been discussed within the framework of
Newtonian cosmology. \citet{do+ry87} showed that for a homogeneous,
isotropic universe, with a Yukawa-type term in the gravitational
interaction, a Newtonian cosmology exists and the equations of motions
are the same as for the ordinary Newtonian case, i.e. the Friedmann
equations. The only difference is that the effective gravitational
constant at large distance, $G_\infty$, enters the equations instead
of the local measured value, $G_\mathrm{N}$. Although no proof has
been produced that such predictions from a Newtonian cosmology can be
used in full relativistic cosmology, this modified FLRW model with an
effective coupling constant provides a motivated framework where
implications of a modified cosmic expansion rate on density
fluctuations can be explored.

We should therefore consider using the standard form of Friedmann
equations, with $G_\infty$ in place of $G_\mathrm{N}$, and the
standard cosmological parameters, with the only change being to
normalize energy densities to a re-scaled cosmological critical
density, $\rho_\mathrm{cr}^\mathrm{Ykw} \equiv 3 H_0^2/\left( 8 \pi
G_\infty
\right)$. Hence, the usual cosmological parameters are related to the new
ones by
\beq
\label{csm1}
\Omega_{i0} = \frac{G_\mathrm{N}}{ G_\infty } \Omega_{i0}^\mathrm{Ykw} =
\frac{\Omega_{i0}^\mathrm{Ykw}}{\alpha_\mathrm{Nwt}}.
\eeq
Within this model, local estimates of the matter density parameter, as
those from cluster galaxy mass and luminosity, should be re-scaled
when being compared to values determined from cosmic distances.

A Yukawa-type potential causes a change in the cosmic expansion rate
of the background. The effect on linear perturbations is easily
understood. Now, also the Hubble drag term in Eq.~(\ref{lin1}) depends
on the effective coupling constant $G_\infty$, whereas the form of the
driving term on the right hand side is not affected. A positive
long-range interaction ($\Delta \alpha>0$) enhances the drag term in
Eq.~(\ref{lin1}), so that growth is suppressed with respect to the non
modified cosmological dynamics. We rewrite the linear perturbation
equation as
\beq
\label{csm2}
\ddot{\delta}_\bfk + 2 H \dot{\delta}_\bfk+
\frac{3}{2} \frac{\Omega_\mathrm{M0}^\mathrm{Ykw} H_0^2}{a^3}
\frac{{\cal Y}}{\alpha_\mathrm{Nwt} } \delta_\bfk = 0,
\eeq
with
\beq
\label{csm3}
H (a) =H_0 \sqrt{\Omega_\mathrm{M0}^\mathrm{Ykw} a^3 +\ol^\mathrm{Ykw}
+
\ok^\mathrm{Ykw}a^2}.
\eeq
As can be seen after comparison with Eq.~(\ref{lin2bis}),
Eq.~(\ref{csm2}) can not be simply obtained by substituting each
cosmological density parameter $\Omega_{i0}$ with the corresponding
$\Omega_{i0}^\mathrm{Ykw}$, since the driving term is suppressed by a
factor $1/\alpha_\mathrm{Nwt}$ due to the new role played by the effective
coupling constant in the background dynamics. Let us consider the
equivalent of an EdS model, i.e a flat model with
$\Omega_\mathrm{M0}^\mathrm{Ykw}=1 $. The perturbation equation can
now be solved in term of hypergeometric functions,
\begin{eqnarray}
\label{csm4}
\delta_\bfk & =& C_1 \left( \frac{a}{a_\lambda}\right)^{-3/2+p_\beta} \\
& {\times} &
\,_2F_1
\left[ \frac{p_\beta}{2} -\frac{5}{4} , \frac{p_\beta}{2},
p_\beta -\frac{1}{4},- \left( \frac{a}{a_\lambda}\right)^2
\right] \nonumber \\
& + &  C_2 \left( \frac{a}{a_\lambda}\right)^{1-p_\beta} \nonumber \\
& {\times} &\,_2F_1
\left[ -\frac{p_\beta}{2}, \frac{5}{4} -\frac{p_\beta}{2} ,
\frac{9}{4} - p_\beta ,- \left( \frac{a}{a_\lambda}\right)^2 \nonumber
\right],
\end{eqnarray}
where $\,_2F_1$ is the hypergeometric function, $C_1$ and $C_2$ are
normalization constants determined by the initial conditions and $p_\beta
\equiv \left( 5 - \sqrt{25 - 24\beta}\right)/4$ with $ \beta \equiv \Delta
\alpha/(1+\Delta \alpha) $. The second term in
Eq.~(\ref{csm4}) is the growing mode which interpolates between
$\delta \propto a^{1-p_\beta}$ at small scales ($a \ll a_\lambda$) and
the usual $\delta \propto a$ at late times and/or large scales. Due to
the modified background, a Yukawa-like contribution with $\Delta
\alpha >0 (<0)$ determines suppression (extra-growth) at small scales,
whereas there is no variation on large scales.

The global effect on the shape of growth function is actually similar
to the case of a non-modified cosmic expansion rate discussed in the
previous sections, the main difference being an overall normalization
factor. The two cases are really indistinguishable when the
differences in the growth index between large and small scale, i.e.
$p_{\Delta \alpha}$ and $p_{\beta}$, are equal. Let us further explore
the near-degeneracy between modified and non-modified cosmic expansion
by considering the EdS model in the two cases. We recall that
the different definitions of critical density,
$\Omega_\mathrm{M0}^\mathrm{Ykw}=1$ or $\Omega_\mathrm{M0}=1 $
correspond to different values of $\rho_\mathrm{M0}$. A value $\Delta
\alpha$ in a non-modified background distorts the power spectrum in
nearly the same way as $\Delta \alpha'= ( 3\Delta \alpha
-4p_{\Delta
\alpha}^2) /\left[4p_{\Delta \alpha}^2 +3(1-\Delta \alpha) \right]$ in a
modified cosmology with approximately the same cutoff proper length.
In Fig.~\ref{GrowthFunction_ykw_mod}, we show the degeneracy in the
shape of the growth function between modified and non-modified cosmic
expansion rate. Differences show themselves only in the normalization
factor, with growth of fluctuation suppressed for enhanced long-range
gravitational interaction, due to the boosted cosmic expansion.

\section{Conclusions}
\label{sec:con}

We have discussed the dynamics of large-scale structure in the
presence of a Yukawa-like contribution to the gravitational potential.
The strength parameter and its cutoff length can introduce peculiar
scale-dependent features in the shape of the power spectrum of density
fluctuations. The main imprint of enhanced gravity is enhanced
large-scale power, in a way that resembles the effect of a lower matter
density. If the extra long-range force couples to all kinds of matter
in the same way, then baryons and CDM evolve together after
recombination. Furthermore, for $\lambda \gs 10~h^{-1}\mathrm{Mpc}$,
the transfer function at the Compton drag epoch is not significantly
affected by Yukawa-like contributions.

With regard to evolution, enhanced gravity acts very similarly to hot
dark matter: both effects yield a scale dependence of the growth
function, even at low redshifts, with a suppression at small scales.
Leakage of gravity on a proper scale of $\lambda \gs
10~h^{-1}\mathrm{Mpc}$ could nearly completely erase the signatures of
non-zero neutrino mass on the shape of power spectrum, making the
detection harder and softening the upper limit on the mass.

A modified cosmic expansion rate should not affect the scale-dependent
distortion of the power spectrum shape due to Yukawa-like
contributions. Whereas the exact dependence of the shape on both
strength and cutoff length is slightly dependent on back-reaction from
modified expanding background, a detection of deviation from the
inverse-square law should be independent of the model for the cosmic
expansion rate.

The degeneracies discussed suggest caution when determining limits on
deviation from the inverse-square law from galaxy clustering using
sharp priors. At the same time, constraints on cosmological
parameters, such as $\om$ and $f_\nu$, could be affected by modified
gravity physics. Full information from acoustic oscillations and
evolution could disentangle some degeneracies. Furthermore, even if
many imprints due to modified gravity are degenerate with other
effects with respect to the shape, they are not with respect to
normalization. While hot dark matter suppresses galaxy clustering on
small scales, enhanced gravity boosts perturbation growth on large
scales. At the same time, in a cosmic expansion enhanced by enhanced
gravity, perturbations are strongly suppressed and the standard growth
rate is achieved only on large scales.

We have discussed deviations from the inverse-square law in the
standard framework of adiabatic initial conditions but it remains an
open question whether a modified gravitational force could make room
for isocurvature initial conditions \citep{pee02}.

Following a phenomenological approach, a Yukawa-like modification to the potential can be 
introduced without being derived from some specific gravity theory. Although such an empirical analysis 
can be enough to investigate the effects of deviations from the inverse-square gravity law, 
some further theoretical considerations can be of interest. Variation of the gravitational 
constant can arise naturally in scalar-tensor theories. Whereas a massive scalar field can 
produce a Yukawa-like modification \citep{zh+ne94}, the existence of a massless scalar field 
coupled to the tensor field of Einstein gravity can implement a time varying $G$. 
The Jordan-Brans-Dicke theory is the original and simplest extended theory of gravity 
which leads to time variations in $G$ \citep{br+di61} and has provided a suitable framework 
for investigation of the effects of a changing $G$ on structure formation \citep{lid+al98,nag+al02}. 
The main effect on the matter power spectrum is connected to a shift to a higher redshift of the 
matter-radiation equality which on turn leads to a shift of the maximum of the matter power spectrum and to more small-scale power. 

A time-varying gravitational constant has an effect quite distinct from that of
a spatially varying gravitational constant, as considered in this paper in the form of a Yukawa-like contribution to the potential. 
In fact, due to the lack of a characteristic scale during matter domination, a time-varying $G$ 
will alter the growth rate but it will not change the shape of the spectrum, unlike a spatially-varying $G$, which naturally 
introduces a cutoff length scale. This main difference characterises potential parameter degeneracies. 
As we have seen, a Yukawa-like modification can act like hot dark matter; on the other hand, the shift 
in the power spectrum due to a time-varying $G$ has a potential degeneracy with increasing the number of 
massless species. Effects of either spatial or temporal variations in $G$ could be better probed via their effects 
on the overall normalization of the power spectrum and its evolution, although biasing is a major source of uncertainty. 

Constraints on a Yukawa-like contribution to the potential have been found in the laboratory and analyses of planetary motion 
in the solar system can extend this to $\lambda \sim 1$~AU \citep{ald+al03}. Such bounds on $\Delta \alpha$ can be very tight, but they 
do not apply to scales beyond the solar system and give no information for $\lambda \gs 10$~Mpc, i.e. the length scales we have been 
considering with regards to structure formation. On these scales, cosmology offers the only probe of modified gravity models.
At present, we have seen that the near-degeneracy between modified gravity and
changes in the cosmological parameters means that exotic gravitational models remain hard to exclude.
However, such degeneracies can be broken with good enough data \citep[e.g.][]{SawCar} and we
can look forward to increasingly stringent tests of Einstein gravity on the largest scales.

\section*{Acknowledgements}
MS thanks the Institute for Astronomy, University of Edinburgh, for
hospitality and financial support during his visit.
JAP was supported by a PPARC Senior Research Fellowship. This work has been
partially supported by the MIUR grant COFIN2004020323.

\setlength{\bibhang}{2.0em}

\end{document}